\documentclass[aps,prl,twocolumn,10pt,superscriptaddress,floatfix]{revtex4-2}

\usepackage{amsmath,amssymb,amsfonts,bm}
\usepackage{graphicx}
\usepackage[colorlinks=false]{hyperref} % FIX: APS-friendly links
\graphicspath{{Figures/}{./}}

\begin{document}

\title{Pressure-Free Surface-Induced Flow by Geometric Rectification}

\author{Zheng Li}
\affiliation{Independent Researcher, Seattle, WA, USA}
\date{\today}

\begin{abstract}
        Pressure-driven flow collapses when confined ($u\propto r^{2}$).
        Asymmetry rectifies surface activity (exchange or slip gradients) into axial flux at $\Delta P=0$ despite zero net exchange.
        Lorentz reciprocity yields a projection law: throughput is the inner product of source with a geometry kernel.
        Signatures include inverted ``narrower-is-faster'' scaling ($u\propto r^{-1}$), leading-order viscosity independence, length amplification ($Q\propto L$), and linear superposition, defining surface-induced flow as a pressure-free Stokes-transport mode from microfluidics to physiology.
\end{abstract}

\maketitle

%%%%%%%%%%%%%%%% %INTRODUCTION %%%%%%%%%%%%%%%%

In a tube of radius $r$, Poiseuille flow gives a mean speed $u\propto r^{2}$ at fixed pressure drop per length~\cite{HappelBrenner1983}, so pressure-driven transport fades under confinement.
Yet experiments show sustained transport even at $\Delta P=0$, including a counterintuitive narrower-is-faster trend in hydrogel-lined tunnels~\cite{Li2020}.
These observations point to \emph{surface-induced} flow: interfacial activity can drive axial throughput without an imposed pressure drop.
In the lubrication/Stokes setting, surface activity enters the 1D flux balance through two boundary channels-wall-normal exchange (transpiration across the boundary) and wall-tangential actuation (slip or tangential stress, e.g.\ electro-/diffusio-osmotic slip driven by solute gradients~\cite{Anderson1989}).
With activity fixed per unit wall area, the driving scales with perimeter ($\propto r$) while the response is averaged over cross-sectional area ($\propto r^{2}$), giving the geometric inverse-radius trend $u\sim r^{-1}$ in slender channels.

Scaling alone does not explain directed transport.
A key question is rectification: if surface activity has no global bias, with neither an imposed pressure drop nor any net mass flux across the boundary, why does net throughput emerge at all?
Kinematic reversibility forbids net flux for mirror-symmetric combinations of activity and geometry, but permits finite throughput once symmetry is broken.
Geometric asymmetry can therefore rectify unbiased surface activity into directed current, akin to Brownian ratchets~\cite{Marchesoni2009,Reimann2002}.
To isolate rectification in its cleanest form, we take wall-normal mass exchange with zero net exchange as a minimal representative of surface-induced forcing.
Classical suction and injection flows in porous channels are typically studied under imposed pressure drops~\cite{Berman1953,Sellars1955,Yuan1956,Oxarango2004,Galowin1974}; here we instead isolate the pressure-free response to boundary activity alone, using exchange as a tractable starting point before extending the same projection structure to wall-tangential actuation.

Here we derive a projection law for the pressure-free volumetric throughput.
The novelty of this formulation lies not in any specific surface mechanism—such as suction, injection, or slip—but in the identification of a geometry-only sensitivity kernel, with surface activity reducing to an effective one-dimensional forcing whose inner product with this kernel determines the axial throughput.
This formulation unifies normal exchange and tangential actuation: exchange specifies an effective source directly, while tangential slip enters as a boundary-driven flux that can be recast as an equivalent source in the 1D balance.
The projection law yields four signatures of surface-induced flow: inverse-radius size dependence ($u\sim r^{-1}$), viscosity independence, length amplification (throughput $\propto L$ at fixed activity density), and linear superposition.
More broadly, pressure-free throughput arises whenever surface activity has a nonzero projection onto the geometric sensitivity kernel.
The result follows from Lorentz reciprocity in the lubrication limit~\cite{MichelinLauga2015,MasoudStone2019}, and it naturally encompasses exchange-driven physiological settings, such as Starling-type transvascular exchange driving axial perfusion in microvessels~\cite{Starling1896,Levick2010,Li2023}.

\begin{figure}[t]
        \centering
        \includegraphics[width=\columnwidth]{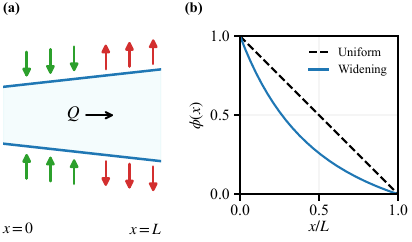}
        \caption{Surface-induced transport model (illustrated here for wall-normal exchange).
                (a) Schematic of confined channel with varying width.
                Walls exhibit distributed injection (green, $s>0$) and removal (red, $s<0$).
                (b) Geometry-dependent sensitivity profile $\phi(x)$ (downstream fraction of total hydraulic resistance), weighting local wall exchange contribution to net axial flux.
                In a widening channel (solid), $\phi(x)$ is biased toward the narrow end, so upstream exchange contributes more strongly than downstream exchange.
        }
        \label{fig:model}
\end{figure}

Consider incompressible Newtonian flow in a slender channel of length $L$ and varying cross-sectional area $A(x)$.
Surface activity is encoded by an \emph{effective} volumetric source per unit length $s(x)$ (positive adds to the axial throughput, negative removes it).
For permeable walls, $s(x)$ coincides with true wall-normal exchange; for impermeable walls driven by tangential interfacial slip, the same role is played by an equivalent source $s(x)=-\mathrm{d}q_b/\mathrm{d}x$ (defined below).
We take the slender limit $\varepsilon \equiv a/L \ll 1$, where $a$ is the transverse scale (e.g., radius $r$).
In this regime, leading lubrication applies.
Define flux $q(x)=A(x)u(x)$, where $u(x)$ is the cross-section-averaged axial velocity (a scalar). We denote the full velocity field by $\mathbf{v}$.
Mass conservation gives
\begin{equation}
        \frac{\mathrm{d}q}{\mathrm{d}x}=s(x),
        \label{eq:mass}
\end{equation}
and we focus on the zero-net-exchange condition,
\begin{equation}
        \int_0^L s(x)\,\mathrm{d}x = 0,
        \label{eq:zero_net_source}
\end{equation}
so channel neither accumulates nor depletes fluid globally (schematic in Fig.~\ref{fig:model}).

\begin{figure}[t]
        \centering
        \includegraphics[width=0.75\columnwidth]{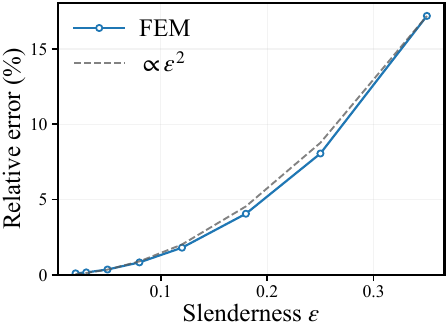}
        \caption{Validation of lubrication theory against 2D finite-element simulations.
                Relative error between 2D FEM and 1D lubrication prediction versus slenderness $\varepsilon$, consistent with expected $O(\varepsilon^2)$ accuracy.
        }
        \label{fig:validation}
\end{figure}

Integrating Eq.~\eqref{eq:mass} gives
\begin{equation}
        q(x) = Q + \int_0^x s(x')\,\mathrm{d}x',
        \label{eq:q_solution}
\end{equation}
where $Q \equiv q(0)$ is the constant throughput.
To determine $Q$, we use the lubrication relation
\begin{equation}
        q(x) = -\kappa(x)\,\frac{\mathrm{d}P}{\mathrm{d}x},
        \label{eq:lubrication}
\end{equation}
where the local hydraulic conductance $\kappa(x)>0$ is set by the channel geometry and viscosity $\mu$; for a circular tube, $\kappa(x)=\pi r(x)^4/(8\mu)$.
We validated this 1D reduction against 2D Stokes FEM, finding relative error scales as $O(\varepsilon^2)$ (Fig.~\ref{fig:validation}).

Across the channel, we define the pressure drop as $\Delta P \equiv P(0)-P(L)$.

Using $\mathrm{d}P/\mathrm{d}x=-q/\kappa$ gives
\begin{equation}
        \Delta P=\int_0^L \frac{q(x)}{\kappa(x)}\,\mathrm{d}x.
        \label{eq:dP}
\end{equation}
Define total hydraulic resistance
\begin{equation}
        R=\int_0^L \frac{\mathrm{d}x}{\kappa(x)}.
        \label{eq:R_def}
\end{equation}

Substituting Eq.~\eqref{eq:q_solution} into Eq.~\eqref{eq:dP} and solving yields
$Q=Q_{\mathrm{press}}+Q_{\mathrm{ex}}$, where $Q_{\mathrm{press}}\equiv \Delta P/R$ and
\begin{equation}
        Q_{\mathrm{ex}} \equiv -\frac{1}{R}\int_{0}^{L}\frac{1}{\kappa(x)} \left[\int_{0}^{x}s(x')\,\mathrm{d}x'\right]\mathrm{d}x.
        \label{eq:Qex_def}
\end{equation}

This shows net throughput is the linear superposition of pressure-driven and surface-induced contributions.
Setting $\Delta P=0$ leaves $Q=Q_{\mathrm{ex}}$, so transport arises solely from exchange.
To reveal rectification, we swap the order of integration in $Q_{\mathrm{ex}}$ and obtain a projection form~\cite{SuppMat}:

\begin{equation}
        Q = -\int_0^L s(x)\,\phi(x)\,\mathrm{d}x,
        \label{eq:adjoint_result}
\end{equation}
with sensitivity profile
\begin{equation}
        \phi(x)=\frac{1}{R}\int_x^L \frac{\mathrm{d}\xi}{\kappa(\xi)}.
        \label{eq:phi_def}
\end{equation}

\begin{figure}[t]
        \centering
        \includegraphics[width=\columnwidth]{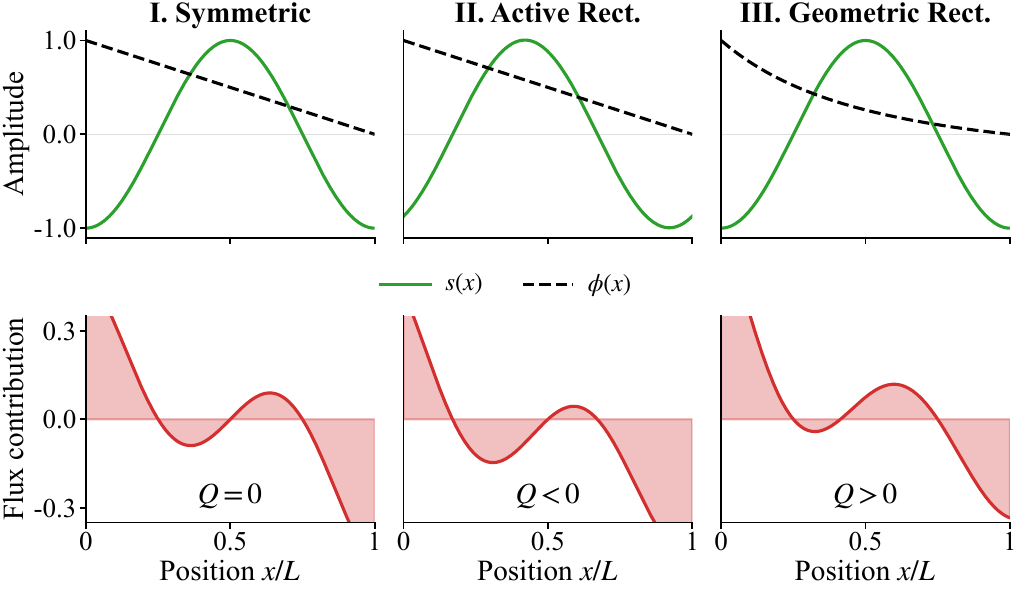}
        \caption{Mechanisms of hydrodynamic rectification.
                (I) Symmetric null: in a uniform channel, mirror-symmetric exchange profiles cancel ($Q=0$).
                (II) Exchange rectification: broken symmetry in $s(x)$ yields $Q\neq 0$ even for uniform channels.
                (III) Geometric rectification: broken symmetry in $\kappa(x)$ yields $Q\neq 0$ even for symmetric $s(x)$, via asymmetric sensitivity $\phi(x)$.}
        \label{fig:symmetry}
\end{figure}

Here $\phi(x)$ is geometric: dimensionless, bounded ($0\le\phi\le1$), fraction of total hydraulic resistance downstream of $x$.
Equation~\eqref{eq:adjoint_result} exposes a rectification structure: in linear Stokes flow, pressure-free throughput from surface activity is inner product of an effective 1D source with a geometry-only Green's function $\phi(x)$.
Eq.~\eqref{eq:adjoint_result} is the 1D lubrication limit of the Lorentz reciprocal theorem, projecting boundary velocities onto the stress field of an auxiliary unit-throughput problem~\cite{MichelinLauga2015,MasoudStone2019}.
Our derivation renders this theorem into a closed-form kernel $\phi(x)$ involving only resistance distribution, giving an explicit tool for pump design.
Geometric asymmetry makes $\phi(x)$ nonuniform, so injection and suction with zero net exchange contribute unequally, yielding rectification.
Transport is symmetry controlled.
If the configuration is mirror symmetric about $L/2$, with $\kappa(x)=\kappa(L-x)$ and $s(x)=s(L-x)$, then $\phi(x)+\phi(L-x)=1$ and, using Eq.~\eqref{eq:zero_net_source}, contributions cancel ($Q=0$).
Directed transport thus requires broken mirror symmetry in exchange, geometry, or their combination (Fig.~\ref{fig:symmetry}).

This separation extends beyond wall-normal exchange.
Any surface activity that enters the 1D flux balance as a boundary-driven term is rectified by the same kernel $\phi(x)$.
Let surface activity add a boundary-driven flux $q_b(x)$ to the axial throughput,
\begin{equation}
        q(x)=-\kappa(x)\,\frac{\mathrm{d}
                P}{\mathrm{d}x}+q_b(x).
\end{equation}
For impermeable walls, incompressibility gives $\mathrm{d}q/\mathrm{d}x=0$, which recasts the boundary-driven term as an equivalent source in the pressure-driven balance,
\begin{equation}
        s(x)= -\frac{\mathrm{d}q_b}{\mathrm{d}x}.
\end{equation}
Note that $q_b(x)$ is defined up to an additive constant; only its derivative enters the equivalent source and thus the projection onto $\phi(x)$.
Therefore the same projection onto $\phi(x)$ holds.
For spatially varying wall-tangential actuation with effective surface velocity $u_s(x)$, $q_b(x)=A(x)\,u_s(x)$, and the resulting $s(x)$ is rectified by $\phi(x)$.

Surface-induced rectification efficiency is viscosity independent at leading order because $\mu$ enters both $\kappa(x)$ and $R$ and cancels in $\phi(x)$.
However, geometric transfer efficiency differs from the total flux.
While $\phi(x)$ is purely geometric, the effective driving strength $s(x)$ can depend on pressure and viscosity through the underlying boundary physics (e.g., Starling-type exchange with $s\propto P/\mu$, or slip laws that set $q_b$).
Equation~\eqref{eq:adjoint_result} thus separates rectification, through $\phi(x)$, from driving rheology encoded in $s(x)$.
The projection law relies only on 1D mass balance and linear hydraulic response, and remains applicable when $s(x)$ is determined self-consistently, provided the resulting closure preserves linearity.

We next ask how large rectified throughput can be under zero net exchange.
Using $\int_0^L s\,\mathrm{d}x=0$, Eq.~\eqref{eq:adjoint_result} becomes $Q=-\int_0^L s(x)\,[\phi(x)-\bar{\phi}]\,\mathrm{d}x$, where $\bar{\phi}\equiv L^{-1}\int_0^L \phi(x)\,\mathrm{d}x$.
Assume a pointwise bound on the 1D source, $|s(x)|\le S_0$.
Then
\begin{equation}
        |Q|\le S_0\int_0^L |\phi(x)-\bar{\phi}|\,\mathrm{d}x
        \le \frac{1}{2}\,S_0 L ,
        \label{eq:scaling_Q}
\end{equation}
where the last inequality uses $0\le\phi\le1$ (see Supplemental Material~\cite{SuppMat}).
Thus bounded zero-mean activity gives $|Q|\lesssim S_0L$.
For a circular tube, the effective 1D source per unit length can be written as $s(x)=2\pi r\,J(x;r)$ with $J$ the wall activity per unit area (e.g.\ wall-normal exchange speed).
Then $S_0\sim 2\pi r\,J_0(r)$ and $u=Q/(\pi r^2)\sim (L/r)\,J_0(r)$, recovering $u\propto r^{-1}$ when the areal activity scale $J_0$ is radius-independent.

\begin{figure}[t]
        \centering
        \includegraphics[width=\columnwidth]{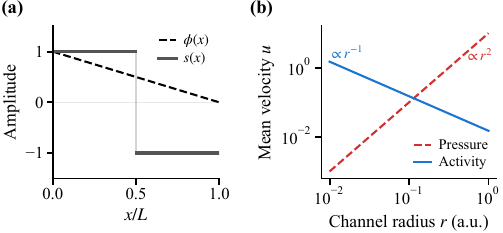}
        \caption{Optimization and scaling.
                (a) In a uniform channel, throughput is maximized by opposing injection and removal across the midpoint.
                (b) Poiseuille flow weakens in confinement ($u \propto r^2$).
                Surface-induced flow has a geometric baseline $u_{\rm ex}\propto r^{-1}$ when $J_0$ is $r$ independent (for wall-normal exchange, $J_0=v_w$), and dominates below $r_c$ defined by $\Gamma=1$.
        }
        \label{fig:scaling}
\end{figure}

For a uniform channel $\phi(x)=1-x/L$, the first inequality in Eq.~\eqref{eq:scaling_Q} is sharp: it is attained by a single switch at the midpoint, as shown in Fig.~\ref{fig:scaling}(a),
\begin{equation}
        s_{\rm opt}(x)=S_0\,\mathrm{sgn}(L/2-x),
\end{equation}
giving $|Q|_{\max}=S_0L/4$.
This linear bound is the origin of length amplification.

Length amplification is not free.
At $\Delta P=0$, throughput is sustained by an internal pressure field.
With $Q\sim S_0L$, the pressure scale is $P_{\rm int}\sim Q R$ with $R=\int_0^L \kappa^{-1}\,\mathrm{d}x$.
For nearly uniform $\kappa$, $R\sim L/\kappa$, hence $P_{\rm int}\sim S_0L^2/\kappa$.
Thus $Q\propto L$ cannot persist in a single long conduit, because the required internal pressure excursions grow with length.
Biological conduits often cap pressure excursions by segmentation.
Collecting lymphatics are partitioned by one-way valves into lymphangions, setting an effective segment length scale~\cite{LevickBook2010}.
In plants, phloem sieve plates partition sieve tubes into sieve elements and provide a major contribution to hydraulic resistance~\cite{Jensen2012SievePlates}.

Energetically, the surface-induced flow is not free.
Sustained transport requires a maintained driving imbalance.
An end-to-end pressure drop $\Delta P$ supplies this imbalance at the ends, whereas the term $s(x)$ represents a distributed drive supplied through surface interactions along the wall.
Setting $\Delta P=0$ therefore removes end-driving but not the need for power: as long as $s(x)\neq 0$, energy must be supplied and is dissipated by viscosity.
Surface-induced flow and pressure-driven flow therefore define two distinct driving limits, whose crossover sets the relevant scale.

To compare surface-induced flow with ordinary pressure-driven flow, we define the speed
ratio
\begin{equation}
        \Gamma \equiv \frac{|u_{\rm ex}|}{|u_{\rm press}|},
        \label{eq:Gamma_def}
\end{equation}
so the crossover occurs at $\Gamma\sim 1$, as summarized in Fig.~\ref{fig:scaling} (b).
Geometry enters twice: through the kernel $\phi(x)$ and through the reduction of wall activity to the one-dimensional source $s(x)$.
More generally,
\begin{equation}
        \frac{\mathrm{d}\phi}{\mathrm{d}x}=-\frac{1}{R\,\kappa(x)},
\end{equation}
so exchange is weighted most strongly where $\kappa(x)$ is smallest.
For a tube, represent wall activity by an areal flux scale $J_0(r)$ so that $S_0\sim (2\pi r)\,J_0(r)$ (for wall-normal exchange, $J_0=v_w$).
With $Q_{\rm ex}\sim S_0L$,
\begin{equation}
        u_{\rm ex}\sim \frac{Q_{\rm ex}}{\pi r^2}\sim \frac{J_0(r)\,L}{r},
\end{equation}
so $u_{\rm ex}\propto r^{-1}$ is the geometric baseline when $J_0$ is $r$ independent.
With $\kappa\sim r^4/\mu$ and $u_{\rm press}\sim \Delta P\,r^2/(\mu L)$,
\begin{equation}
        \Gamma \sim \frac{\mu L^2}{\Delta P}\,\frac{J_0(r)}{r^3},
\end{equation}
which separates geometry ($r^{-3}$) from source strength ($J_0$).
The condition $\Gamma\sim 1$ gives $J_0(r_c)/r_c^3\sim \Delta P/(\mu L^2)$, reducing to $r_c\sim (\mu J_0 L^2/\Delta P)^{1/3}$ when $J_0$ is $r$ independent.
This decomposition gives a clean reading of confinement trends: $r^{-1}$ is the rectification baseline, while any confinement dependence in $J_0(r)$ steepens or weakens it without changing the projection structure.
Experiments~\cite{Li2020} reporting exponents steeper than $r^{-1}$ are consistent with an $r$-dependent $J_0(r)$ on top of this geometric baseline.

Though derived in 1D, this formulation is exact at leading order for Stokes flows in slender 2D/3D conduits.
Integrating $\nabla\cdot\mathbf{v}=0$ over a cross-section yields Eq.~\eqref{eq:mass}, where $s(x)$ is the effective surface-induced source per unit length.
For true wall-normal exchange, $s(x)$ equals the net wall-normal flux per unit length.
In 3D this is a line integral around the cross-section boundary,
\begin{equation}
        s(x)=\oint_{\partial A(x)} \mathbf{v}\cdot\mathbf{n}\,\mathrm{d}\ell,
        \label{eq:universality}
\end{equation}
where $\mathbf{n}$ is the unit normal pointing into the fluid.
For impermeable walls with boundary-driven tangential slip, the same 1D balance is obtained by recasting the boundary-driven contribution $q_b(x)$ as $s(x)=-\mathrm{d}q_b/\mathrm{d}x$.
The corresponding 2D reduction is obtained per unit depth by evaluating normal flux through walls.
Linearity guarantees the projection structure onto $\phi(x)$ carries over to arbitrary shapes via $\kappa(x)$.

In biology, walls can be compliant.
Under quasi-static deformation, the leading-order effect is a renormalization of the local hydraulic response $\kappa(x)$.
Rectification persists provided the deformation preserves the axial asymmetry of the normalized conductance profile.

We close by summarizing the main implication of the projection law.
At $\Delta P=0$, the surface-induced flow is set by the projection $Q=-\int_0^L s(x)\,\phi(x)\,\mathrm{d}x$; it vanishes when this projection is zero and is finite otherwise.
Because it relies only on 1D mass balance and linear hydraulic response, the same kernel extends at leading order to slender 2D/3D conduits.

These considerations also suggest a microfluidic testbed.
In synthetic conduits, one can program an effective $s(x)$ using patterned permeability, electroosmotic wall slip, osmotic coatings, or distributed injection and removal.
Matching $s(x)$ to the sensitivity profile $\phi(x)$ then becomes a direct design rule to generate and control throughput at $\Delta P=0$, and to explore confinement regimes where near-wall effects can reduce hydraulic load.

The same viewpoint connects naturally to microvascular hemodynamics.
In microvessels, transvascular exchange is ubiquitous~\cite{Starling1896,MichelCurry1999,Woodcock2012,Michel2020}, and vessel walls exhibit metabolically maintained surface activity, including electrokinetic effects~\cite{Weinbaum2007,Tarbell2014,Pollack2013FourthPhase} and osmotic interfacial stresses~\cite{Anderson1989}.
Together these processes generate a persistent, distributed source term \(s(x)\).
Because axial geometric asymmetry is unavoidable in vascular networks, \(s(x)\) is generically not orthogonal to the sensitivity profile \(\phi(x)\).
Therefore the surface-induced contribution to axial transport is generically nonzero: at \(\Delta P=0\) it yields a finite throughput \(Q=-\int_0^L s(x)\phi(x)\,\mathrm{d}x\), and under physiological perfusion it persists as an additive distributed drive.

These ingredients have two hydrodynamic roles in different regimes.
First, geometric asymmetry rectifies sustained surface activity through the projection law, producing a baseline throughput even at \(\Delta P=0\) \((Q\propto S_0L)\).
Second, under strong confinement, for example when red blood cells restrict flow to thin near-wall plasma layers, the same distributed activity acts as effective lubrication, reducing the hydraulic load by modifying the hydraulic response where near-wall transport dominates.
Together, these two roles identify biological microvessels as natural realizations of surface-driven, geometry-rectified transport.

Poiseuille’s law describes transport driven by bulk pressure gradients, whereas our theory captures the complementary surface-driven regime that dominates under tight confinement.
Confinement suppresses pressure-driven flow, $u \propto r^{2}$, but enhances surface-induced flow, $u \propto r^{-1}$, making this Stokes-transport mode increasingly relevant at high surface-to-volume ratios.

\begin{acknowledgments}
        The author is grateful to G.~H.~Pollack for proposing the original question that initiated this line of work.
        Although the theoretical and mathematical modelling developed here differs from his preferred perspective, this work would not exist without his pioneering insights.
        The author also thanks W.~Kaminsky, K.~Böhringer, R.~J.~Wilkes, and H.~Lai for valuable discussions.
        Finally, the author is especially grateful to A.~Li and R.~Hua for their unwavering support.
\end{acknowledgments}

\section*{Data Availability Statement}
The data that support the findings of this study are available within the article and its Supplemental Material.
\bibliographystyle{apsrev4-2}
\bibliography{References/references}

\end{document}

\title{Supplemental Material:\\
        Pressure-Free Surface-Induced Flow by Geometric Rectification}
\author{Zheng Li}
\affiliation{Independent Researcher, Seattle, WA, USA}
\date{\today}

\maketitle

Section~S1 derives the 1D reduction and yields the $\phi(x)$ identity for $Q$.

Section~S2 validates the 1D reduction against finite-element (FEM) simulations.

Section~S3 summarizes symmetry constraints, rectification design criteria, and conservative bounds.

Section~S4 develops scaling estimates, including the inverted-confinement scaling.

Section~S5 details the agarose microtunnel experiments used in the thick/thin comparison.

\section{1D lubrication theory and the geometric projection law}

\subsection{Physical setup, definitions, and assumptions}
We consider an incompressible Newtonian fluid in a slender channel of length \(L\), with cross-sectional area \(A(x)\) varying smoothly along the axial coordinate \(x\in[0,L]\).
At the cross-sectionally averaged (1D) level, wall-normal activity is modeled by an effective volumetric source per unit axial length \(s(x)\) (dimensions \([\mathrm{L}^2\mathrm{T}^{-1}]\)): \(s(x)>0\) denotes net injection into the bulk and \(s(x)<0\) net removal.

We define:
\begin{itemize}
        \item \(u(x)\): cross-sectionally averaged axial velocity,
        \item \(q(x)\equiv A(x)\,u(x)\): volumetric flux,
        \item \(Q \equiv q(0)\): inlet flux (global throughput).
\end{itemize}

Our assumptions are:
\begin{itemize}
        \item Newtonian, incompressible flow: constant density and viscosity \(\mu\); the bulk satisfies \(\nabla\cdot\mathbf{v}=0\).
        \item Slender geometry (lubrication regime): \(\varepsilon \equiv a/L \ll 1\) (or equivalently \(h/L\ll 1\), depending on notation) and Reynolds number is sufficiently low that inertia is negligible.
              Consequently, the leading-order pressure is uniform across each cross-section, \(P\approx P(x)\).
        \item 1D reduction with distributed exchange: velocities are cross-sectionally averaged, and all wall activity enters through the axial source density \(s(x)\).
\end{itemize}

It is convenient to separate \emph{net} exchange from \emph{spatially balanced} exchange by decomposing
\begin{equation}
        s(x)=\bar s+\delta s(x),
        \qquad
        \bar s\equiv \frac{1}{L}\int_0^L s(x)\,dx,
        \qquad
        \int_0^L \delta s(x)\,dx=0.
        \label{eq:supp_s_decomp}
\end{equation}
Here \(\bar s\) measures the mean (net) injection/removal per unit length, while \(\delta s(x)\) captures exchange that is locally nonzero but has zero net.
From 1D continuity, a nonzero mean implies a net flux difference across the channel,
\begin{equation}
        q(L)-q(0)=\int_0^L s(x)\,dx = L\bar s.
        \label{eq:supp_net_flux_diff}
\end{equation}
When we wish to isolate \emph{surface-induced rectification} (throughput generation without net addition/removal), we set \(\bar s=0\), equivalently
\begin{equation}
        \int_0^L s(x)\,dx = 0 \quad \Longrightarrow \quad q(L)=q(0)=Q.
        \label{eq:supp_zero_mass}
\end{equation}
This zero-mean condition is a \emph{special case} rather than a requirement: the projection identity derived below remains valid for general \(s(x)\).

\subsection{Governing 1D lubrication model}
Under the lubrication approximation (slender geometry, low Reynolds number), the leading-order pressure is uniform across each cross-section and the flux obeys a Darcy-like relation
\begin{equation}
        q(x) = -\kappa(x)\,\frac{dP}{dx},
        \label{eq:supp_darcy}
\end{equation}
where \(\kappa(x)>0\) is the local hydraulic conductance. Wall exchange enters through 1D mass conservation,
\begin{equation}
        \frac{dq}{dx}=s(x).
        \label{eq:supp_continuity}
\end{equation}
Integrating \eqref{eq:supp_continuity} from \(0\) to \(x\) gives the local flux distribution
\begin{equation}
        q(x)=Q+\int_{0}^{x}s(x')\,dx'.
        \label{eq:supp_q_from_s}
\end{equation}

\subsection{Sensitivity kernel \texorpdfstring{$\phi(x)$}{phi(x)} and projection identity}
Rearranging \eqref{eq:supp_darcy} yields
\begin{equation}
        \frac{dP}{dx}=-\frac{q(x)}{\kappa(x)}.
        \label{eq:supp_Pprime}
\end{equation}
Integrating from \(0\) to \(L\) and defining the imposed pressure drop \(\Delta P \equiv P(0)-P(L)\) gives the global pressure balance
\begin{equation}
        \Delta P=\int_{0}^{L}\frac{q(x)}{\kappa(x)}\,dx.
        \label{eq:supp_DeltaP_q}
\end{equation}
Substituting \eqref{eq:supp_q_from_s} into \eqref{eq:supp_DeltaP_q} produces
\begin{equation}
        \Delta P
        =\int_{0}^{L}\frac{1}{\kappa(x)}
        \left[
                Q+\int_{0}^{x}s(x')\,dx'
                \right]dx.
        \label{eq:supp_DeltaP_sub}
\end{equation}
Define the total hydraulic resistance
\begin{equation}
        R\equiv \int_{0}^{L}\frac{dx}{\kappa(x)}.
        \label{eq:supp_def_R}
\end{equation}
The source-dependent contribution in \eqref{eq:supp_DeltaP_sub} is a double integral over \(0\le x'\le x\le L\).
Swapping the order of integration (Fubini) gives
\begin{equation}
        \int_{0}^{L}\frac{1}{\kappa(x)}
        \left[\int_{0}^{x}s(x')\,dx'\right]dx
        =
        \int_{0}^{L}s(x')\left[\int_{x'}^{L}\frac{dx}{\kappa(x)}\right]dx'.
        \label{eq:supp_swap}
\end{equation}
This motivates the dimensionless geometric sensitivity kernel
\begin{equation}
        \phi(x)\equiv \frac{1}{R}\int_{x}^{L}\frac{d\xi}{\kappa(\xi)},
        \label{eq:supp_def_phi}
\end{equation}
which satisfies \(\phi(0)=1\) and \(\phi(L)=0\), and decreases monotonically with \(x\).

Using \eqref{eq:supp_swap}--\eqref{eq:supp_def_phi} in \eqref{eq:supp_DeltaP_sub} (and relabeling dummy variables) yields \[ \Delta P = Q R + R\int_{0}^{L}s(x)\phi(x)\,dx.
\]
Solving for \(Q\) gives the superposition form
\begin{equation}
        Q=\frac{\Delta P}{R}-\int_{0}^{L}s(x)\phi(x)\,dx.
        \label{eq:supp_Q_general}
\end{equation}
In the self-pumping case \(\Delta P=0\), this reduces to the projection law
\begin{equation}
        Q=-\int_{0}^{L}s(x)\phi(x)\,dx.
        \label{eq:supp_Q_projection}
\end{equation}

Using the decomposition \eqref{eq:supp_s_decomp} in the self-pumping case \eqref{eq:supp_Q_projection} yields an explicit split,
\begin{equation}
        Q =-\bar s\int_0^L \phi(x)\,dx \;-\; \int_0^L \delta s(x)\,\phi(x)\,dx \;\equiv\; Q_{\mathrm{net}}+Q_{\mathrm{bal}}.
        \label{eq:supp_Q_split}
\end{equation}
The first term \(Q_{\mathrm{net}}\) represents the throughflow associated with nonzero net exchange, while \(Q_{\mathrm{bal}}\) is the rectified contribution from the zero-mean component.
Accordingly, the symmetry cancellations and conservative bounds derived below pertain to \(Q_{\mathrm{bal}}\) (i.e., the case \(\bar s=0\)).

\subsection{Tangential slip as an equivalent source term}
Suppose tangential wall actuation contributes an additive boundary-driven flux \(q_b(x)\), so that
\begin{equation}
        q(x) = -\kappa(x)\,\frac{dP}{dx} + q_b(x).
        \label{eq:supp_q_with_qb}
\end{equation}
For impermeable walls, \(dq/dx=0\) and hence \(q(x)\equiv Q\).
Then \[ \frac{dP}{dx} = -\frac{Q-q_b(x)}{\kappa(x)}.
\]
Integrating from \(0\) to \(L\) gives the global relation
\begin{equation}
        \Delta P
        = \int_0^L \frac{Q-q_b(x)}{\kappa(x)}\,dx
        = Q R - \int_0^L \frac{q_b(x)}{\kappa(x)}\,dx,
\end{equation}
so that
\begin{equation}
        Q=\frac{\Delta P}{R}+\frac{1}{R}\int_0^L \frac{q_b(x)}{\kappa(x)}\,dx.
        \label{eq:supp_Q_slip_general}
\end{equation}
In particular, for \(\Delta P=0\),
\begin{equation}
        Q=\frac{1}{R}\int_0^L \frac{q_b(x)}{\kappa(x)}\,dx.
        \label{eq:supp_Q_slip_DP0}
\end{equation}

Equivalently, using \(\phi'(x)=-(1/R)\kappa(x)^{-1}\) and \(\phi(L)=0\), one may rewrite \eqref{eq:supp_Q_slip_DP0} as
\begin{equation}
        Q = q_b(0) + \int_0^L \frac{dq_b}{dx}\,\phi(x)\,dx,
        \label{eq:supp_Q_slip_phi}
\end{equation}
which reduces to \(Q=\int_0^L q_b'(x)\phi(x)\,dx\) if \(q_b(0)=0\) by convention.

\section{Finite-element validation}
This section validates the 1D projection law against 2D steady incompressible Stokes flow in a slender planar channel \(\Omega=\{(x,y):0<x<L,\ |y|<h(x)\}\), with linear taper \(h(x)=h_{\rm in}+(h_{\rm out}-h_{\rm in})x/L\) and \(h_{\rm out}/h_{\rm in}=2\) (we set \(L=1\) without loss of generality).
We imposed a distributed exchange profile \(s(x)=\cos(\pi x/L)\) by prescribing a purely normal wall velocity on \(y=\pm h(x)\):
\begin{equation}
        \mathbf{v}\cdot \mathbf{n} \;=\; \frac{s(x)}{2\sqrt{1+h'(x)^2}},
\end{equation}
Here $\mathbf{n}$ points into the fluid, consistent with the main text,
so that the net inward flux through both walls over an interval \(dx\) equals \(s(x)\,dx\).
No velocity was prescribed at \(x=0\) and \(x=L\) (natural traction conditions).

To match the pressure-free (self-pumping) setting, we enforced a zero mean pressure drop, \(\langle p\rangle_{x=0}=\langle p\rangle_{x=L}\), and fixed the pressure gauge by setting \(p\) at one point (here \(p\) is the FEM pressure field, the same physical pressure as \(P\) in the 1D model).
We discretized using Taylor--Hood elements (P2/P1) on mapped triangular meshes and verified mesh independence (refining the mesh changed \(Q_{\rm FEM}\) by less than \(1\%\)).

From each FEM solution we computed the section flux \(q(x)=\int_{-h(x)}^{h(x)} v_x(x,y)\,dy\) at several interior probe locations and estimated \(Q_{\rm FEM}=q(x)-\int_0^x s(\xi)\,d\xi\) (averaged over probes).
We compared \(Q_{\rm FEM}\) with the 1D prediction \(Q_{1\mathrm{D}}=-\int_0^L s(x)\phi(x)\,dx\), using \(\phi\) computed from planar lubrication scaling \(\kappa(x)\propto h(x)^3\).

\section{Bounds on Rectified Throughput}
This section derives the bound on rectified throughput used in the main text.
We start from the projection law (Sec.~S1),
\begin{equation}
        Q = -\int_0^L s(x)\,\phi(x)\,\mathrm{d}x.
        \label{eq:Q_projection_SI}
\end{equation}

\noindent\textbf{Net exchange versus rectified (zero-mean) exchange.}
The projection law \eqref{eq:Q_projection_SI} holds for general activity profiles \(s(x)\), including cases with nonzero net exchange.
It is convenient to decompose \(s(x)=\bar s+\delta s(x)\) with \(\bar s=L^{-1}\int_0^L s\,dx\) and \(\int_0^L\delta s\,dx=0\).
Then \eqref{eq:Q_projection_SI} implies the exact split
\begin{equation}
        Q =-\bar s\int_0^L \phi(x)\,dx \;-\; \int_0^L \delta s(x)\,\phi(x)\,dx \;\equiv\; Q_{\mathrm{net}}+Q_{\mathrm{bal}}.
        \label{eq:Q_split_net_bal_SI}
\end{equation}
The first term \(Q_{\mathrm{net}}\) is a net-throughflow contribution associated with \(\bar s\neq 0\).
The second term \(Q_{\mathrm{bal}}\) is the \emph{rectified} contribution from the zero-mean component of the activity.
Accordingly, the symmetry cancellation arguments and the conservative bound derived below apply to \(Q_{\mathrm{bal}}\), i.e.\ to the surface-induced (zero-net-exchange) setting
\begin{equation}
        \int_0^L s(x)\,\mathrm{d}x=0.
        \label{eq:zero_mean_s}
\end{equation}

\noindent Throughout, the kernel satisfies the pointwise bounds \(0\le \phi(x)\le 1\).

Introduce the spatial mean of the kernel,
\begin{equation}
        \bar{\phi}\equiv \frac{1}{L}\int_0^L \phi(x)\,\mathrm{d}x .
        \label{eq:phi_bar_def}
\end{equation}
Using \eqref{eq:zero_mean_s}, we may subtract any constant from \(\phi\) inside the overlap without changing \(Q\), since \[ \int_0^L s(x)\,\bar\phi\,\mathrm{d}x=\bar\phi\int_0^L s(x)\,\mathrm{d}x=0.
\]
Hence \eqref{eq:Q_projection_SI} is equivalently written in the centered form
\begin{equation}
        Q = -\int_0^L s(x)\,\big[\phi(x)-\bar{\phi}\big]\,\mathrm{d}x .
        \label{eq:Q_centered}
\end{equation}
Intuitively, in a zero-net-exchange setting only the \emph{spatial variation} of \(\phi\) can contribute; its absolute offset is irrelevant.

Assume a pointwise activity bound \(|s(x)|\le S_0\).
Taking absolute values in \eqref{eq:Q_centered} and using \(|\int g|\le \int |g|\) gives
\begin{align}
        |Q|
         & = \left|\int_0^L s(x)\,[\phi(x)-\bar\phi]\,\mathrm{d}x\right| \nonumber \\
         & \le \int_0^L |s(x)|\,|\phi(x)-\bar\phi|\,\mathrm{d}x \nonumber          \\
         & \le S_0\int_0^L |\phi(x)-\bar\phi|\,\mathrm{d}x ,
        \label{eq:Q_bound_step1}
\end{align}
i.e.
\[
        |Q|\le S_0\int_0^L \big|\phi(x)-\bar{\phi}\big|\,\mathrm{d}x .
\]

We now bound the centered \(L^1\) deviation for any function \(f:[0,L]\to[0,1]\).
Let \(\bar f=\frac{1}{L}\int_0^L f\).
Then
\begin{equation}
        \int_0^L |f(x)-\bar f|\,\mathrm{d}x \le 2\,\bar f\,(1-\bar f)\,L \le \frac{L}{2}.
        \label{eq:L1_mean_bound_strong}
\end{equation}
A physical way to see this is to ask: for fixed mean \(\bar f\) and bounded amplitude \(0\le f\le 1\), how do we maximize the total deviation from the mean?
Values of \(f\) close to \(\bar f\) contribute little to \(|f-\bar f|\), so to increase \(\int|f-\bar f|\) one should push \(f\) as far from \(\bar f\) as allowed, i.e.\ toward the extremal values \(0\) and \(1\).
More concretely, if \(f\) takes an intermediate value in \((0,1)\) on a set of nonzero measure, one can shift part of that set slightly downward and another part slightly upward (keeping the mean fixed) and thereby increase \(|f-\bar f|\) pointwise; iterating this ``sharpening'' procedure drives \(f\) to a two-level (bang--bang) profile without decreasing \(\int|f-\bar f|\).
Thus the maximum is attained by a two-level function taking only values \(0\) and \(1\).
For such a function, the fraction of the interval where \(f=1\) equals \(\bar f\), and a direct computation yields \[ \int_0^L |f-\bar f|\,\mathrm{d}x = \bar f L(1-\bar f) + (1-\bar f)L\bar f = 2\bar f(1-\bar f)L \le \frac{L}{2}, \] with the worst case at \(\bar f=\tfrac12\).

Applying \eqref{eq:L1_mean_bound_strong} to \(f=\phi\) and combining with \eqref{eq:Q_bound_step1} gives the conservative bound
\begin{equation}
        |Q|\le \frac{1}{2}\,S_0\,L ,
        \label{eq:Q_bound_final}
\end{equation}
which is the inequality invoked in the main text.

For a uniform channel \(\phi(x)=1-x/L\), one has \(\bar\phi=\tfrac12\) and \[ \int_0^L \left|\phi-\bar\phi\right|\mathrm{d}x =\int_0^L\left|\frac12-\frac{x}{L}\right|\mathrm{d}x =2\int_0^{L/2}\left(\frac12-\frac{x}{L}\right)\mathrm{d}x =\frac{L}{4}.
\]
Thus \(|Q|_{\max}=S_0L/4\), attained (up to sign) by choosing \(s(x)\) to align with the sign of the centered kernel,
\(s_{\rm opt}(x)=S_0\,\mathrm{sgn}(L/2-x)\).

\section{Scaling-law analysis}

This section unpacks the confinement scaling and the crossover metric $\Gamma$ used in the main text, with special emphasis on the fact that geometry enters twice: (i) through the projection kernel $\phi(x)$, and (ii) through the reduction of a wall-defined activity to an effective one-dimensional forcing.
As shown in Sec.~S1, wall-normal exchange and tangential slip can be treated on the same footing by introducing an effective source.
To match the main text, we denote this effective source by \(s(x)\).
\begin{equation}
        s(x)\equiv s_{\rm w}(x)-\frac{\mathrm{d}q_b}{\mathrm{d}x},
        \label{eq:seff_def}
\end{equation}
where $s_{\rm w}(x)$ is the wall-normal exchange source (if present) and $q_b(x)$ is the boundary-driven flux induced by slip.
With the convention $q_b(0)=0$ (see Sec.~S1 and \eqref{eq:supp_Q_slip_phi}), the pressure-free throughput has the schematic scaling
\begin{equation}
        |Q| \sim \left|\int_0^L s(x)\,\phi(x)\,\mathrm{d}x \right| ,
\end{equation}
so confinement trends separate cleanly into a purely geometric part (through $\phi$) and a mechanism-dependent part contained in $s$.

\subsection{From wall activity to the effective 1D source}

In the 1D lubrication description (Sec.~S1), surface activity enters through an effective signed source term $s(x)$ with units of volume per time per axial length.
For pure wall-normal exchange (no slip), $q_b\equiv 0$ and hence $s(x)=s_{\rm w}(x)$.

We use a circular tube as the running example throughout this section.
For wall-normal exchange, it is convenient to introduce an areal exchange-speed (or areal flux) field $J(x;r)$ with units of speed, defined so that the net volumetric injection per unit axial length is the wall integral of $J$.
For a tube of radius $r(x)$ this gives
\begin{equation}
        s(x)=\oint_{\partial A}
        J\,\mathrm{d}\ell \;\approx\; 2\pi r(x)\,J(x;r).
\end{equation}
Other cross-sections differ only by an order-one geometric prefactor in the reduction from a wall field to a 1D source.

We summarize the magnitude of the wall activity by a representative areal scale $J_0(r)$ (e.g. an rms or peak amplitude in $x$),
so that the corresponding 1D source scale satisfies
\begin{equation}
        S_0 \sim 2\pi r\,J_0(r).
        \label{eq:S0_from_J0}
\end{equation}
This relation is where additional confinement dependence may enter through physics: even when the surface mechanism sets an areal activity scale $J_0$, that scale itself may depend on $r$ (see Sec.~S5).
Similarly, any slip-induced contribution encapsulated in $q_b$ (hence in $s$ via \eqref{eq:seff_def}) can introduce further $r$-dependence.

\subsection{Geometric kernel and weighting}
The projection law from Sec.~S1 can be written schematically as
\begin{equation}
        |Q| \sim \left|\int_0^L s(x)\,\phi(x)\,\mathrm{d}x \right| ,
\end{equation}
where $\phi(x)$ is a purely geometric sensitivity kernel satisfying $\phi(0)=1$, $\phi(L)=0$, and $0\le \phi\le 1$.
In the lubrication derivation (Sec.~S1) the kernel is controlled by the local hydraulic conductance $\kappa(x)$ via
\begin{equation}
        \frac{\mathrm{d}\phi}{\mathrm{d}x}=-\frac{1}{R\,\kappa(x)}, \qquad
        R\equiv\int_0^L \frac{\mathrm{d}\xi}{\kappa(\xi)} .
        \label{eq:dphi_dx_scaling}
\end{equation}
Thus surface activity is weighted most strongly where $\kappa(x)$ is smallest (largest local resistance).
For a tube, the conductance scales as
\begin{equation}
        \kappa(x)\sim \frac{r(x)^4}{\mu},
\end{equation}
where $\mu$ is the dynamic viscosity, up to an order-one geometric factor.

\subsection{Geometric scaling baseline}
We now extract the confinement trend of the exchange-driven mean speed (the case $s=s_{\rm w}$).
The main-text bound (proved in Sec.~S3) implies that for bounded, zero-mean activity,
\begin{equation}
        |Q_{\rm ex}| \lesssim S_0\,L .
        \label{eq:Qex_scaling}
\end{equation}
Combining \eqref{eq:Qex_scaling} with \eqref{eq:S0_from_J0} gives
\begin{equation}
        |Q_{\rm ex}|\;\sim\;(2\pi r)\,J_0(r)\,L .
        \label{eq:Qex_from_J0}
\end{equation}
Dividing by cross-sectional area $A=\pi r^2$ yields the mean exchange-driven speed
\begin{equation}
        u_{\rm ex}\sim \frac{Q_{\rm ex}}{\pi r^2}\sim \frac{J_0(r)\,L}{r}.
        \label{eq:uex_inv_scaling}
\end{equation}
Equation \eqref{eq:uex_inv_scaling} makes the decomposition explicit: the factor $L/r$ is purely geometric, while any additional confinement dependence is contained in the areal activity scale $J_0(r)$.
In particular, when $J_0$ is independent of $r$, the geometric baseline is the inverted confinement scaling $u_{\rm ex}\propto r^{-1}$.

\subsection{Crossover with pressure-driven flow}
For ordinary Poiseuille flow in a tube driven by a pressure drop $\Delta P$ over length $L$,
the mean speed scales as
\begin{equation}
        u_{\rm press}\sim \frac{\Delta P}{\mu}\,\frac{r^2}{L},
        \label{eq:upress_scaling}
\end{equation}
again up to an order-one factor (e.g. $1/8$ for a perfect cylinder).

We first recall the main-text definition (exchange-driven case),
\begin{equation}
        \Gamma \equiv \frac{|u_{\rm ex}|}{|u_{\rm press}|}.
        \label{eq:Gamma_def}
\end{equation}
Using \eqref{eq:uex_inv_scaling} and \eqref{eq:upress_scaling} then gives the main-text scaling
\begin{equation}
        \Gamma \sim \frac{\mu L^2}{\Delta P}\,\frac{J_0(r)}{r^3},
        \label{eq:Gamma_scaling}
\end{equation}
up to order-one geometric prefactors.

More generally, for an arbitrary surface mechanism encoded by $s(x)$, we characterize the effective forcing by a typical amplitude \(S_0(r)\) (e.g. peak or rms in \(x\)).
The conservative bound from Sec.~S3 implies
\begin{equation}
        |Q_{\rm si}|\lesssim S_0\,L ,
        \label{eq:Qsi_scaling}
\end{equation}
where the subscript ``si'' denotes the surface-induced (pressure-free) throughput driven by \(s\).
Dividing by area \(A=\pi r^2\) yields a corresponding mean speed scale
\begin{equation}
        u_{\rm si}\sim \frac{Q_{\rm si}}{\pi r^2}\ \lesssim\ \frac{S_0(r)\,L}{\pi r^2}.
        \label{eq:usi_scaling}
\end{equation}
Defining the generalized ratio
\begin{equation}
        \Gamma_{\rm si} \equiv \frac{|u_{\rm si}|}{|u_{\rm press}|},
\end{equation}
and substituting \eqref{eq:usi_scaling} and \eqref{eq:upress_scaling} gives
\begin{equation}
        \Gamma_{\rm si} \sim \frac{\mu L^2}{\Delta P}\,\frac{S_0(r)}{r^4},
        \label{eq:Gamma_scaling_seff}
\end{equation}
again up to order-one geometric prefactors.
For pure wall-normal exchange (no slip), \(s=s_{\rm w}\) and \(S_0=S_0\sim (2\pi r)J_0(r)\), so \(\Gamma_{\rm si}\) reduces to \(\Gamma\) and \eqref{eq:Gamma_scaling_seff} recovers \eqref{eq:Gamma_scaling}.

The crossover $\Gamma_{\rm si}\sim 1$ therefore corresponds to
\begin{equation}
        \frac{S_0(r_c)}{r_c^4}\sim \frac{\Delta P}{\mu L^2},
\end{equation}
which reduces to $r_c\sim (\mu J_0 L^2/\Delta P)^{1/3}$ when exchange dominates and $J_0$ is $r$ independent.

\subsection{Beyond the geometric baseline}
Equation \eqref{eq:uex_inv_scaling} shows that the exchange-driven rectification baseline is $r^{-1}$ at fixed areal activity scale.
In experiments, however, confinement can modify the activity itself (i.e. $J_0(r)$), for example through external transport limitations, surface kinetics, or permeability of the surrounding medium.
Such effects steepen or weaken the apparent confinement exponent without altering the underlying geometric projection structure.
More generally, any additional confinement dependence entering through the effective forcing amplitude \(S_0(r)\) (including slip contributions via \eqref{eq:seff_def}) will modify the observed scaling without changing the geometric kernel weighting.
The agarose-gel example in Sec.~S5 illustrates one way in which the areal activity scale (and hence \(S_0\)) can acquire additional $r$-dependence.

\section{Re-analysis of published agarose microtunnel data}
In this section, we re-analyze the agarose--microtunnel velocity measurements reported by Li and Pollack \cite{Li2020}.
We use the published velocity measurements reported in Fig.
4(B) of Ref. \cite{Li2020}.
These published data are used here only as an external benchmark supporting the statement that exchange-driven flow can scale more strongly than the purely geometric lower bound $u\propto r^{-1}$.
This section documents the dataset, the analysis choices used here, and how the result connects to the geometric projection framework in Sections~S1--S3.

\subsection{Experimental setup and velocity measurements}
Ref.~\cite{Li2020} performed the material-exchange measurements in a $1.5\%$ (w/w) agarose gel cast and trimmed to a rectangular block (approximately $1\,\mathrm{cm}\times1\,\mathrm{cm}\times0.6\,\mathrm{cm}$ in length $\times$ width $\times$ height) \cite{Li2020}.
During the material-exchange configuration, the top gel surface was exposed to ambient air, so macroscopic evaporation provides a distributed sink that sustains solvent exchange through the tunnel boundary \cite{Li2020}.

A straight tunnel of total length $L=1$\,cm was molded through the gel using a composite needle--fiber insert \cite{Li2020}.
The tunnel consists of two equal-length segments ($L/2$ each): a wide segment formed by a needle (diameter $670\,\mu$m) and a narrow segment formed by an optical fiber, with two fiber diameters studied:
\begin{equation}
        d_{\mathrm{thick}}=255\,\mu\mathrm{m},
        \qquad
        d_{\mathrm{thin}}=145\,\mu\mathrm{m},
        \qquad
        \frac{r_{\mathrm{thick}}}{r_{\mathrm{thin}}}
        =
        \frac{d_{\mathrm{thick}}}{d_{\mathrm{thin}}}
        =1.7586.
\end{equation}

Axial velocities $V(t)$ are taken from Fig.
4(B) of Ref.~\cite{Li2020}, where they were measured by particle image velocimetry (PIV) using tracer microspheres.

At early times in Ref.~\cite{Li2020}, the flow is influenced by a surface--water interaction that generates a tracer-depleted near-wall \emph{Exclusion Zone} (EZ).
The EZ forms and grows over tens of minutes \cite{Li2020}.
To keep this EZ-growth period separate from the sustained evaporation-driven forcing, we follow the same practical split used in the main text and focus on $t\ge 100$\,min.
We call this interval an \emph{exchange-dominated} window, where evaporation-driven material exchange provides the main sustained forcing and is comparatively well characterized.

In the published processing, the reported velocity at each time is averaged over the central part of the tunnel cross-section (roughly the central half of the diameter), i.e.\ the high-speed core \cite{Li2020}.
This deemphasizes the EZ-affected near-wall region and makes the PIV estimate more robust.
Because the same protocol is applied to both the thick and thin cases, it may rescale absolute speeds by an $\mathcal{O}(1)$ profile factor but does not, by itself, introduce a systematic bias in the thick/thin \emph{ratio} computed below.

\small
\setlength{\LTleft}{0pt}
\setlength{\LTright}{0pt}

\begin{longtable}{c c c c}
        \caption{Published velocity time series from Ref.~\cite{Li2020} used in Eqs.~(\ref{eq:Gamma_exp_SI})--(\ref{eq:neff_SI}).
                The analysis window for $\overline{\Gamma}_{\mathrm{tt}}$ and $n_{\mathrm{eff}}$ is $t=100,110,\dots,180$\,min.
        }
        \label{tab:S1_timeseries}                                                  \\

        \hline
        $t$ (min)                                      &
        $V_{\mathrm{thick}}$ ($\mu\mathrm{m\,s^{-1}}$) &
        $V_{\mathrm{thin}}$ ($\mu\mathrm{m\,s^{-1}}$)  &
        $|V_{\mathrm{thick}}|/|V_{\mathrm{thin}}|$                                 \\
        \hline
        \endfirsthead

        \multicolumn{4}{l}{\textbf{Table~\thetable} (continued).}                  \\
        \hline
        $t$ (min)                                      &
        $V_{\mathrm{thick}}$ \Vunit                    &
        $V_{\mathrm{thin}}$ \Vunit                     &
        $|V_{\mathrm{thick}}|/|V_{\mathrm{thin}}|$                                 \\
        \hline
        \endhead

        \hline
        \multicolumn{4}{r}{continued on next page}                                 \\
        \endfoot

        \hline
        \endlastfoot

        0                                              & 3.717  & 13.856  & 0.268  \\
        10                                             & 3.027  & 0.274   & 11.042 \\
        20                                             & 1.440  & -1.134  & 1.270  \\
        30                                             & 0.680  & -3.809  & 0.178  \\
        40                                             & 0.464  & -6.388  & 0.073  \\
        50                                             & 0.434  & -8.690  & 0.050  \\
        60                                             & 0.303  & -10.649 & 0.028  \\
        70                                             & 0.046  & -13.479 & 0.003  \\
        80                                             & -0.344 & -14.724 & 0.023  \\
        90                                             & -0.890 & -16.279 & 0.055  \\
        100                                            & -1.706 & -17.241 & 0.099  \\
        110                                            & -1.982 & -17.864 & 0.111  \\
        120                                            & -2.097 & -17.439 & 0.120  \\
        130                                            & -2.150 & -19.291 & 0.111  \\
        140                                            & -2.236 & -20.442 & 0.109  \\
        150                                            & -2.326 & -20.706 & 0.112  \\
        160                                            & -2.455 & -20.289 & 0.121  \\
        170                                            & -2.563 & -19.916 & 0.129  \\
        180                                            & -2.642 & -22.106 & 0.120  \\

\end{longtable}
\normalsize
\subsection{Thick/thin ratio}

Using the post-100\,min window, we compute the thick/thin ratio based on instantaneous absolute speeds:
\begin{equation}
        \overline{\Gamma}_{\mathrm{tt}}
        \;\equiv\;
        \left\langle \frac{|V_{\mathrm{thick}}(t)|}{|V_{\mathrm{thin}}(t)|} \right\rangle_{t\ge 100\,\mathrm{min}}
        =
        0.115 \pm 0.009,
        \label{eq:Gamma_exp_SI}
\end{equation}
where the uncertainty is the sample standard deviation over the $N=9$ time points $t=100,110,\dots,180$\,min.
Equivalently, the thin tunnel is faster by a factor $\overline{\Gamma}_{\mathrm{tt}}^{-1}\approx 8.7$ in this window.

For convenience, we also summarize the observed radius dependence by defining an effective two-point scaling exponent $n_{\mathrm{eff}}$ via $|V|\propto r^{-n_{\mathrm{eff}}}$ over this radius range:
\begin{equation}
        n_{\mathrm{eff}}
        \;=\;
        -\frac{\ln \overline{\Gamma}_{\mathrm{tt}}}{\ln(r_{\mathrm{thick}}/r_{\mathrm{thin}})}
        \;=\;
        3.84 \pm 0.14.
        \label{eq:neff_SI}
\end{equation}
We emphasize that $n_{\mathrm{eff}}$ is a window-dependent summary statistic rather than a claim of a universal asymptotic power law; nonetheless it robustly supports the main-text point that exchange-driven flow can scale more strongly than the geometric lower bound $u\propto r^{-1}$.

\subsection{Geometry--exchange coupling and geometric focusing}
The published thick/thin comparison is naturally interpreted through the projection framework derived in Sections~S1--S3.
Recall that the net throughput is an \emph{overlap} between the exchange (``activity'') profile and a purely geometric sensitivity kernel,
\begin{equation}
        Q=-\int_0^L s(x)\,\phi(x)\,\mathrm{d}x,
        \label{eq:Q_phi_overlap_SI}
\end{equation}
with $q(x)=Q+\int_0^x s(x')\,\mathrm{d}x'$ and $u(x)=q(x)/A(x)$.
The key point is that $s(x)$ is typically \emph{not} an independent control knob: it is set by a coupling between macroscopic sinks (e.g.\ evaporation from the gel surface) and microscopic transport to the tunnel wall.

\paragraph{Macroscopic balance (order-of-magnitude constraint).}
When evaporation provides the dominant macroscopic sink, the overall exchange \emph{demand/supply} is constrained by the evaporation rate and exposed surface area,
\begin{equation}
        S_{\mathrm{tot}}\equiv\int_{0}^{L} s(x)\,\mathrm{d}x \sim E\,A_{\mathrm{surf}},
        \label{eq:global_balance_SI}
\end{equation}
up to order-one factors associated with external humidity and boundary-layer conditions \cite{Li2020,Deegan1997}.
(As a scaling constraint this does not distinguish net exchange from spatially balanced exchange; it only bounds the integrated exchange magnitude.)
Given such a constraint, geometry can still change \emph{how efficiently} exchange is converted into axial throughput: reducing the narrow radius increases downstream hydraulic resistance (sharpening $\phi$), which increases the weight placed on downstream exchange in (\ref{eq:Q_phi_overlap_SI}) and thus enhances geometric focusing.

\paragraph{External transport can introduce additional $r$-dependence.}
As one illustrative limiting case, if transport through the gel to an approximately planar sink is rate-limiting, classical Laplace/transport results imply that the mean wall-normal exchange speed scales approximately as
\begin{equation}
        v_n(r)\sim \frac{D_g\,\Delta\mu}{r\,\ln(2h/r)},
        \qquad (h\gg r),
        \label{eq:vn_line_to_plane_SI}
\end{equation}
where $h$ is the distance to the macroscopic sink, $D_g$ is an effective gel transport coefficient, and $\Delta\mu$ is a driving contrast \cite{Crank1975,Smythe1950,Jackson1999}.
Thus, even before accounting for any additional $r$-dependence of the driving contrast $\Delta\mu$, external transport alone can contribute a factor close to $v_n\propto r^{-1}$ (up to the weak logarithm).

\subsection{Practical implication: geometry matters twice}
This re-analysis highlights a practical point for interpreting composite-channel measurements: internal geometry sets not only the hydraulic/geometric sensitivity (captured by $\phi$), but can also modulate the effective exchange forcing itself through boundary-mediated transport.
In systems where exchange is sustained by external transport (e.g.\ evaporation through the gel), the observed radius dependence can therefore reflect a combination of (i) internal geometric focusing and (ii) radius-dependent exchange strength.
Varying the external transport pathway (for example by changing the depth $h$ and the gel transport coefficient $D_g$ in (\ref{eq:vn_line_to_plane_SI})) can, in principle, separate these contributions while keeping the internal channel geometry fixed.